\documentclass{iau_FM}
\usepackage{graphicx}

\title[Determining the evolutionary stage of HD163899] %% give here short title %%
{Determining the evolutionary stage of HD163899 on the basis of its oscillation spectrum}

\author[Jakub Ostrowski, Jadwiga Daszy\'nska-Daszkiewicz \& Henryk Cugier]   %% give here short author list %%
{Jakub Ostrowski, Jadwiga Daszy\'nska-Daszkiewicz
 \and Henryk Cugier}

\affiliation{Instytut Astronomiczny, Uniwersytet Wroc{\l}awski, ul. Kopernika 11, 51-622 Wroc{\l}aw, Poland \\ email: {\tt ostrowski@astro.uni.wroc.pl}, {\tt daszynska@astro.uni.wroc.pl}, {\tt cugier@astro.uni.wroc.pl}\\[\affilskip]  }

\pubyear{2015}
\jname{Astronomy in Focus, Volume 1}
\editors{}
\begin{document}

\maketitle

\begin{abstract}
We present the new interpretation of the oscillation spectrum of HD 163899 based on the new determinations of the effective temperature, mass-luminosity ratio and rotational velocity. These new parameters strongly prefer the more massive models than previously considered. Now it is also possible that the star could be in the main sequence stage. Using the oscillation spectrum as a gauge, we intend to establish which stage of evolution corresponds better to HD 163899.

\end{abstract}

\firstsection % if your document starts with a section,
\section{Introduction}

\begin{figure}
\begin{center}
 \includegraphics[clip,width=134mm,angle=0]{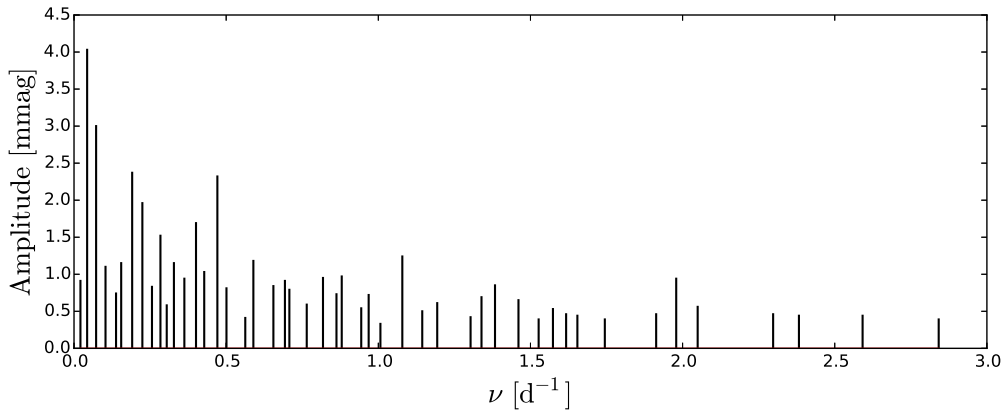}
 \caption{Frequency spectrum of HD 163899 based on the MOST data.}
\label{fig1}
\end{center}
\end{figure}

HD 163899, a prototype of Slowly Pulsating B-type supergiants (SPBsg), has been a subject of interest since \cite[Saio \etal\ (2006)]{saio2006} had found 48 frequencies in its spectrum (Fig.\,\ref{fig1}). These frequencies are in range of $[0.02, 2.85]~\mathrm{d^{-1}}$ and \cite[Saio \etal\ (2006)]{saio2006} attributed them to pulsations in $p$- and $g$-modes. They initially proved to be a challenge for a theory, which did not predict a presence of unstable pulsational modes beyond the main sequence (MS) due to a very efficient damping in the radiative helium core. \cite[Saio \etal\ (2006)]{saio2006} claimed that some pulsational modes can be partially reflected at the intermediate convective zone (ICZ) and hence they do not penetrate the damping region and can be driven in the outer layers. This explanation has been also confirmed by \cite[Ostrowski \& Daszy\'nska-Daszkiewicz (2015)]{ostrowski2015}.

The main problem is establishing the evolutionary status of the star. In this paper we use the new estimate of the stellar parameters based on the high-resolution HARPS spectra. Their values raise again the question (\cite[Godart \etal\ 2008]{godart2008}): is HD 163899 really a supergiant star?

\section{HD 163988}

HD 163899 was classified as B2 Ib/II supergiant by \cite[Klare \& Neckel (1977)]{kn1977} and \cite[Schmidt \& Carruthers(1996)]{sc96}. This spectral type was assigned on the basis of calibrations of photometric indices. There were no precise determinations of the effective temperature, surface gravity or luminosity as well as any information about the rotational velocity of the star. The very rough estimates of $T_{\rm eff}$ and $\log L/L_\odot$ gave the mass in the range [13, 20] $M_\odot$
and the three possibilities for the evolutionary stage: the main sequence, the shell hydrogen burning phase and the core helium burning phase.
The most supported hypothesis was that the star is a blue supergiant in the phase of shell hydrogen burning during the first crossing towards red giant branch
and the core helium burning phase on the blue loop was rejected by \cite[Ostrowski \& Daszy\'nska-Daszkiewicz 2015]{ostrowski2015}.

In order to advance the studies of HD 163899, the proper determinations of its basic physical properties were needed. The high-resolution spectra from the HARPS spectrograph and the atmosphere models of \cite[Lanz \& Hubeny (2007)]{lanz2007} were used to obtain the new parameters: $\log L/M=3.85 \pm 0.05$, $\log T_\mathrm{eff}=23000 \pm 1000$, $\log g = 3.0 \pm 0.15$ and $V \sin i=65 \pm 5$ $\mathrm{km~s^{-1}}$. The star is now slightly hotter and much more luminous
than previously thought (cf. Fig.\,4 in \cite[Saio \etal\ 2006]{saio2006}). As a result it also should be more massive. Such results will have a significant influence on the interpretation of oscillation spectrum of HD 163899.

\section{Theoretical models}

\begin{figure}
\begin{center}
 \includegraphics[clip,width=134mm,angle=0]{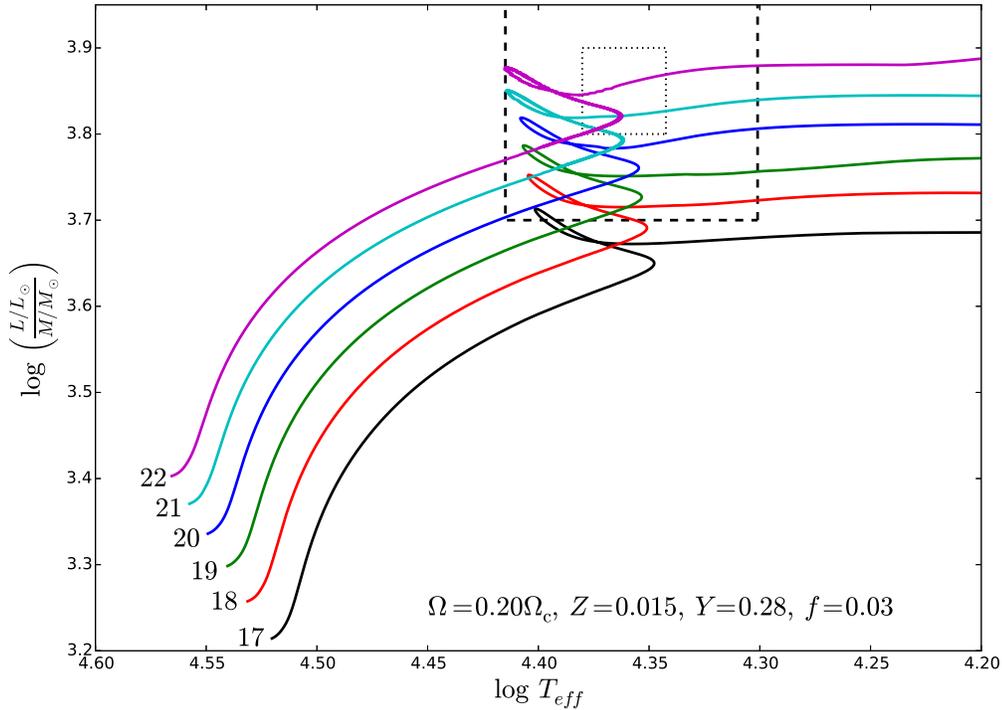}
 \caption{The $\log L/M$ vs. $\log T_\mathrm{eff}$ diagram for models with the initial masses 17 - 22 $M_\odot$, initial rotation of $0.2~\Omega_\mathrm{crit}$, metallicity $Z = 0.015$ and overshooting $f = 0.03$. The inner and outer rectangles show a $1 \sigma$ and $3\sigma$ errors of determined parameters, respectively.}
\label{fig2}
\end{center}
\end{figure}

The presented evolutionary models have been calculated with MESA code (\cite[Paxton \etal\ 2011, 2013, 2015]{paxton11}) and supplemented with pulsation calculations with the nonadiabatic code of \cite[Dziembowski (1977)]{dziembowski1977}. The initial hydrogen abundance of $X=0.70$ and metallicity $Z=0.015$ have been adopted. We used the Ledoux criterion for the convective instability, the OPAL opacity tables (\cite[Iglesias \& Rogers 1996]{opal}) and AGSS09 element mixture (\cite[Asplund \etal\ 2009]{agss09}). Overshooting from the convective core has been calculated with the exponential formula of \cite[Herwig (2000)]{herwig2000} and we treated mass-loss using the prescription of \cite[Vink \etal\ (2001)]{vink01}.We included the rotation assuming $0.2$ of the critical velocity on the zero age main sequence as well as different rotational mixing mechanisms.

The evolutionary tracks for models with initial masses of 17 - 22 $M_\odot$ and overshooting $f=0.03$ are presented in Fig.\,\ref{fig2}. The new values of the stellar parameters have two main consequences for HD 163899. Firstly, it is clear that the star is more massive than it was believed before and has mass higher than 17 $M_\odot$. Secondly, there are evolved main-sequence models in the error box and together with higher mass it increases the likelihood that
HD 163899 is a main-sequence star.

\section{Pulsational models and the HD163899 oscillations}

\begin{figure*}
\begin{center}
 \includegraphics[clip,width=134mm,angle=0]{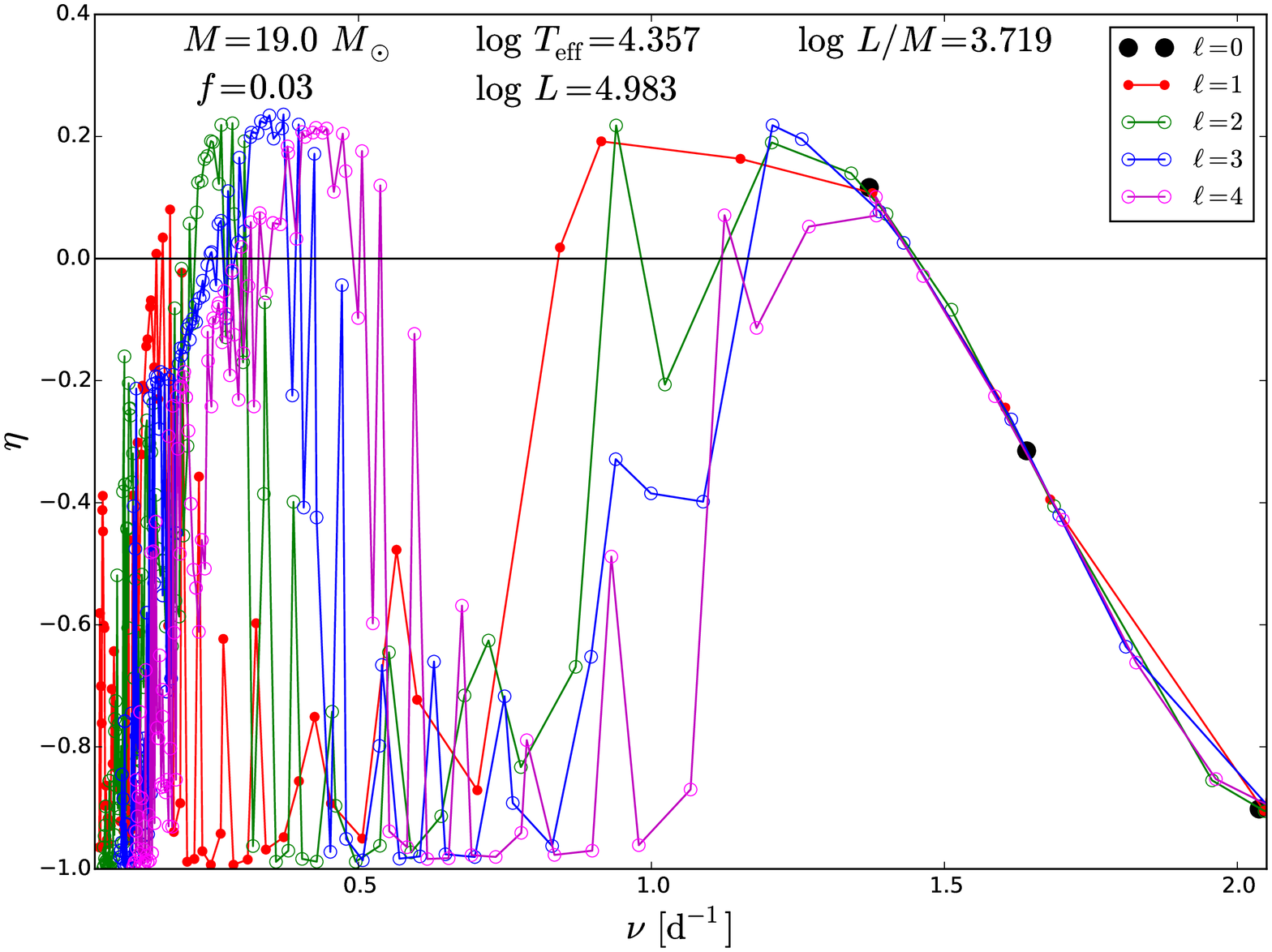}
 \includegraphics[clip,width=134mm,angle=0]{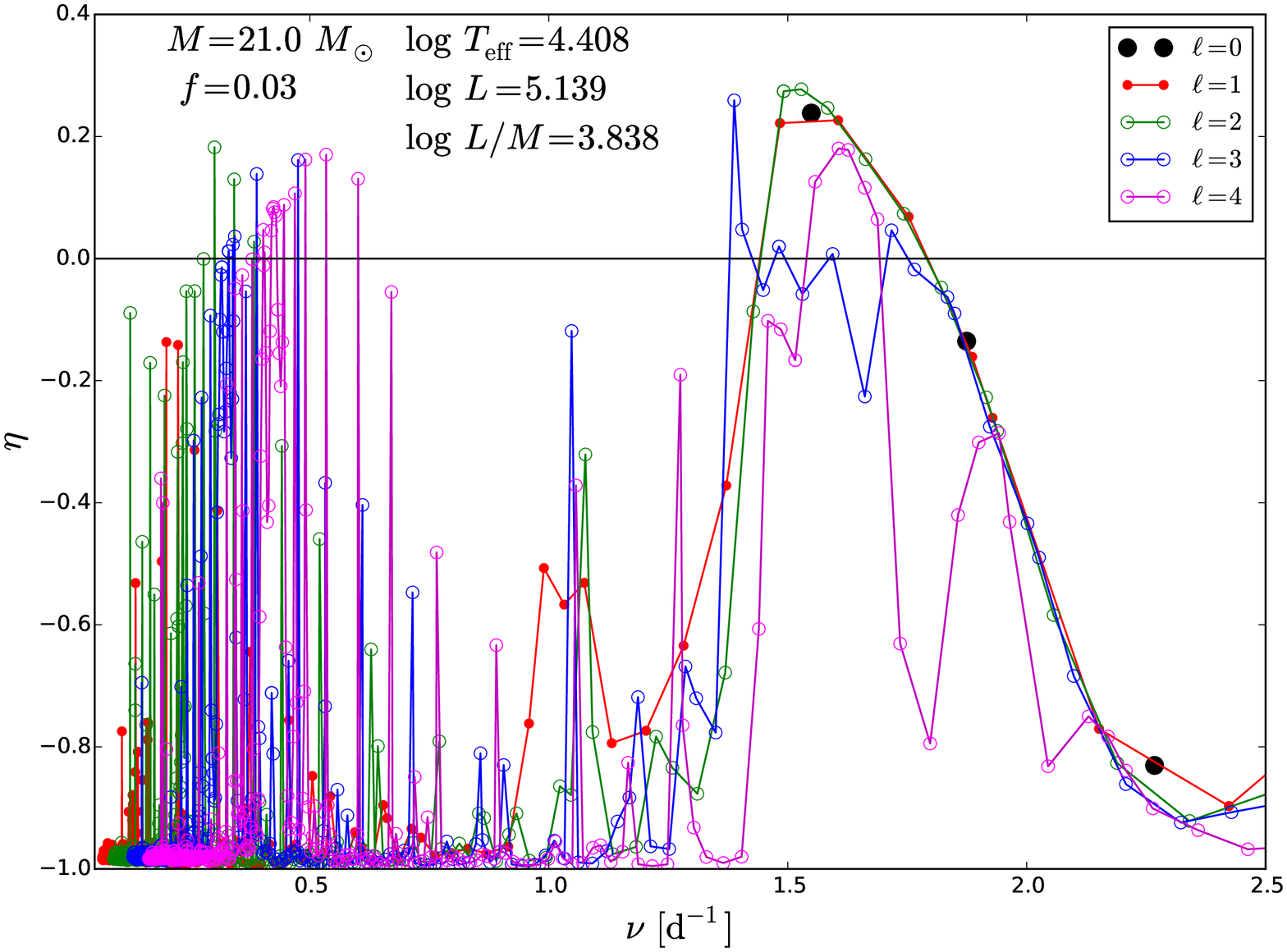}
 \caption{The instability parameter, $\eta$, as a function of frequency, $\nu$, for a main-sequence model with the initial mass $19 M_\odot$ (upper panel) and for a supergiant with the initial mass $21 M_\odot$ (bottom panel). Pulsational modes with the spherical degree up to $\ell = 4$ are presented.}
\label{fig3}
\end{center}
\end{figure*}

\begin{figure*}
\begin{center}
 \includegraphics[clip,width=67mm,angle=0]{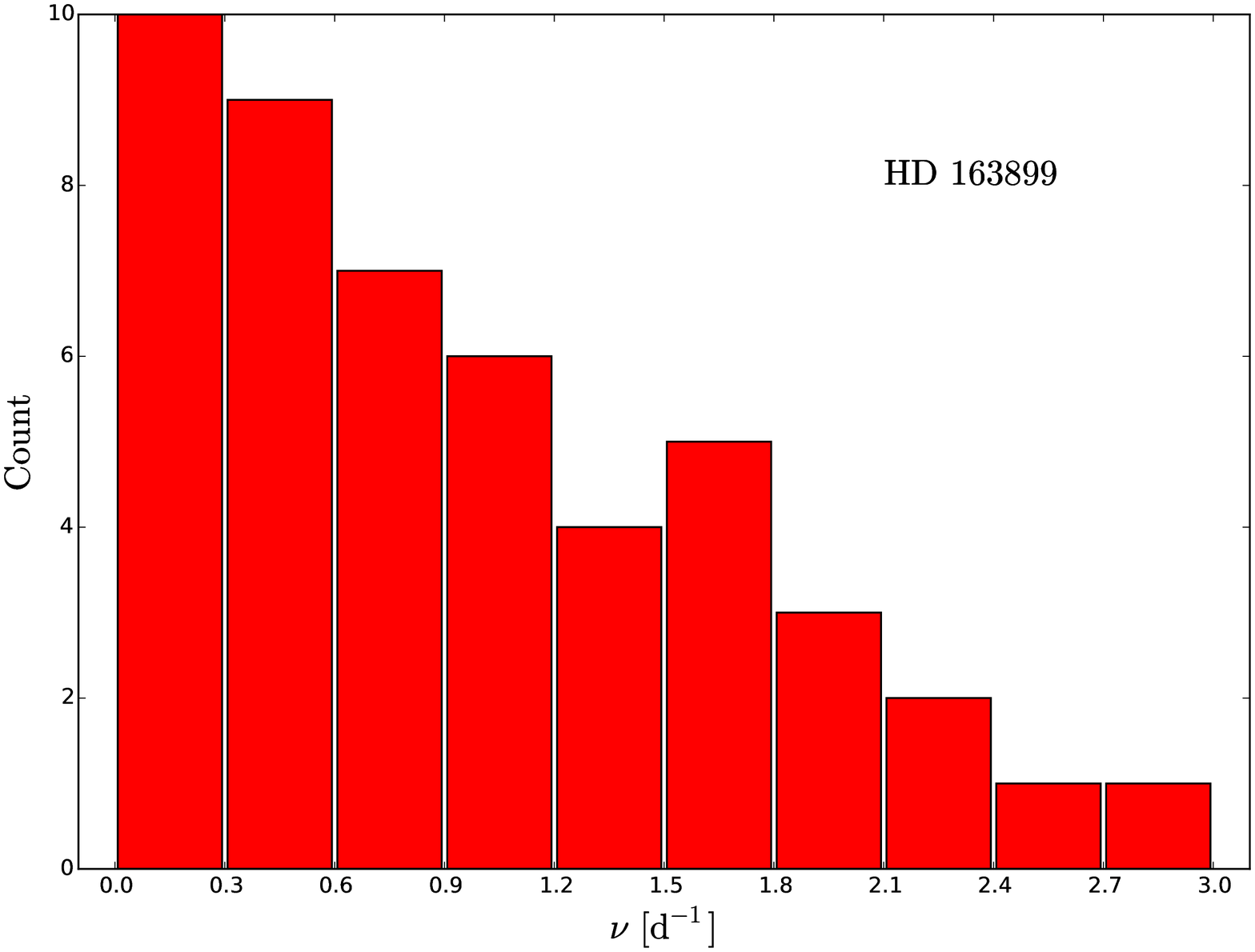}
 \includegraphics[clip,width=67mm,angle=0]{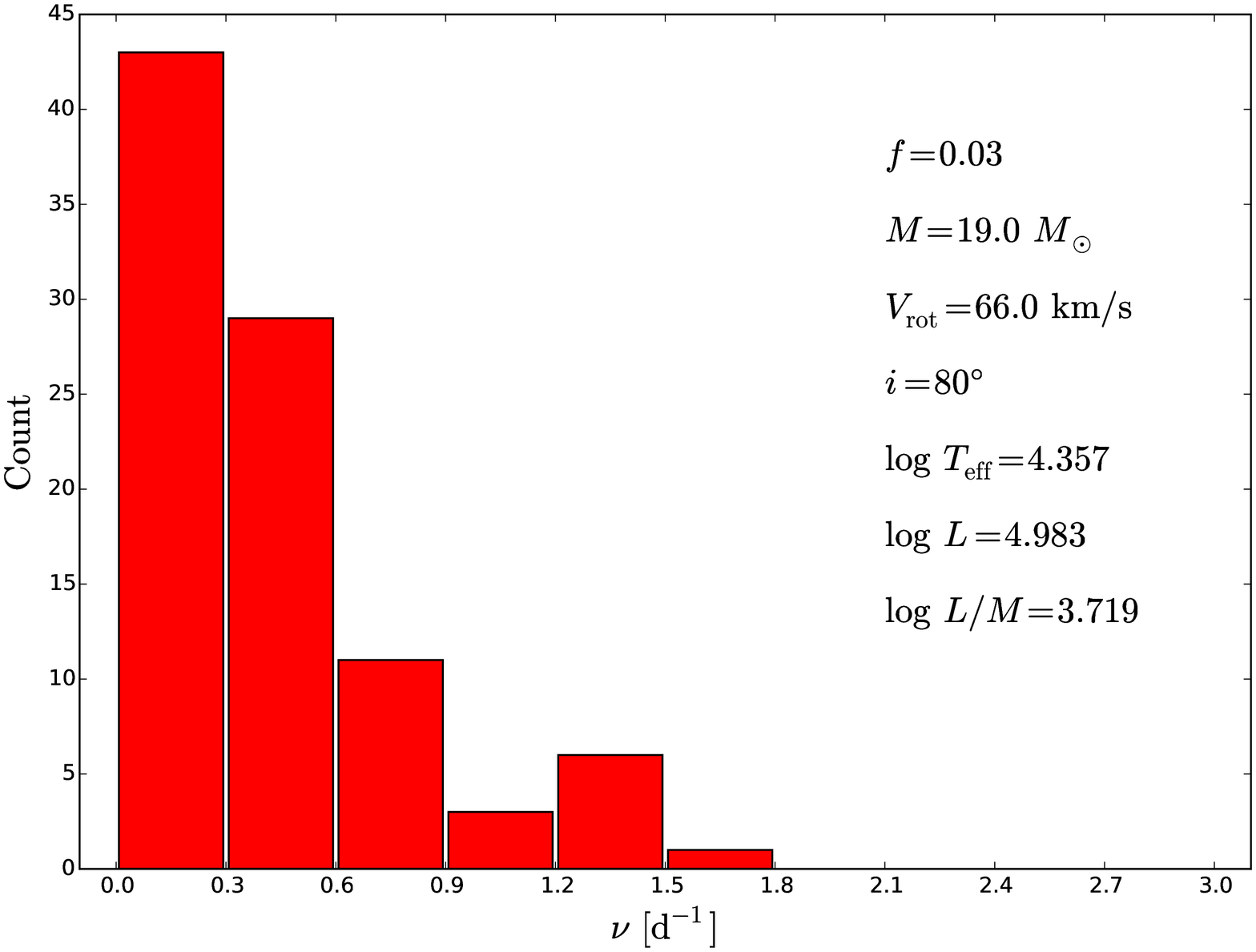}
 \includegraphics[clip,width=67mm,angle=0]{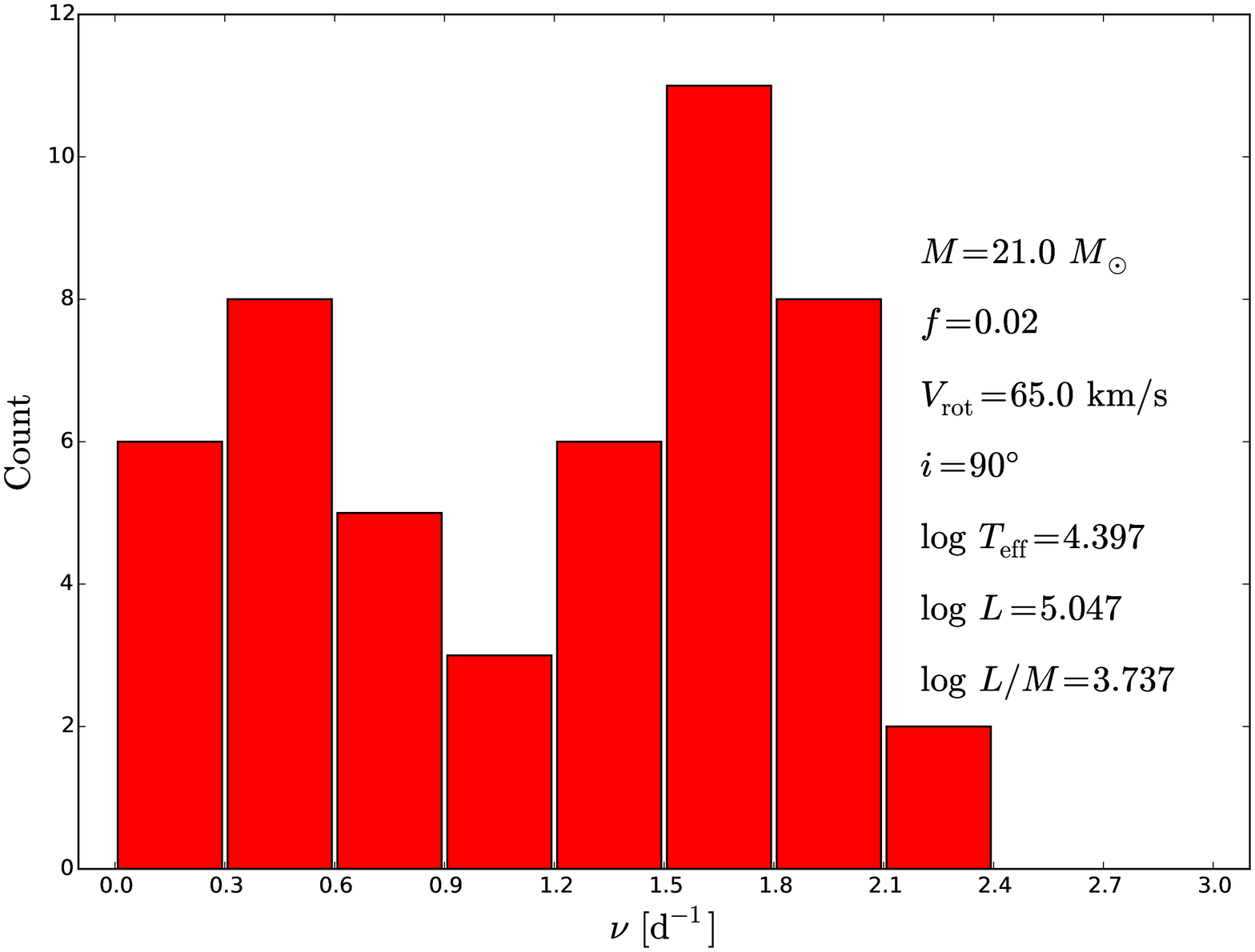}
 \includegraphics[clip,width=67mm,angle=0]{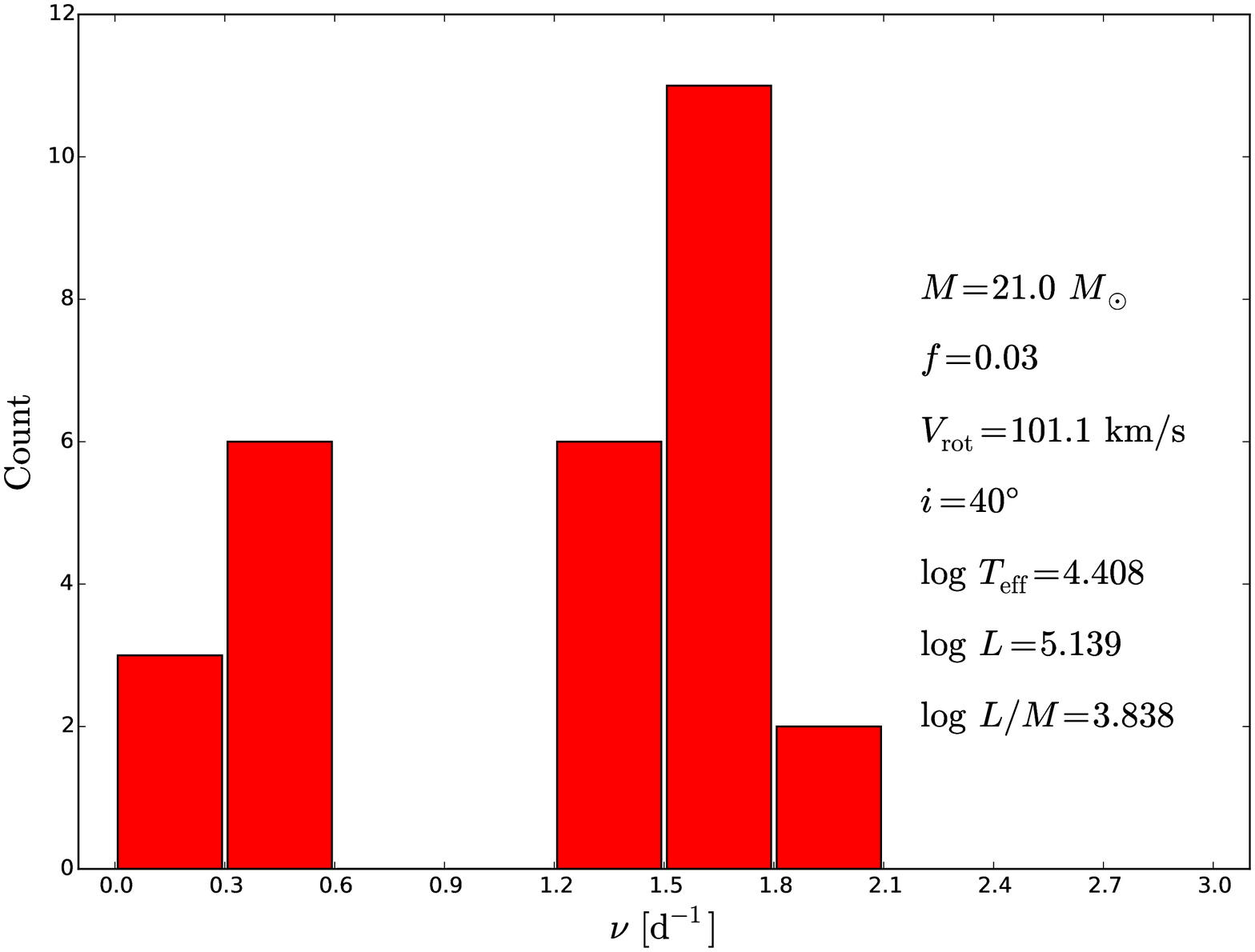}
 \caption{Histogram for the observed frequencies of HD 163899 (the top-left panel). The theoretical histograms were calculated for a main-sequence models with initial masses of $19 M_\odot$ and $21 M_\odot$ (the top-right and bottom-left panels, respectively) and for a supergiant model with initial mass of $21 M_\odot$ (the bottom-left panel). Modes with $\ell=0-4$ and $A_V>0.3$ [mmag] were considered.}
\label{fig4}
\end{center}
\end{figure*}

The instability parameter, $\eta$, tells us whether a pulsation mode is unstable ($\eta > 0$) or not ($\eta < 0$). In Fig.\,\ref{fig3}, we compare the runs of $\eta$ in a function of the oscillation frequency, $\nu$, for two models contained in the error box: a main-sequence (the top panel) and a supergiant model (the bottom panel). Modes with the spherical degrees $\ell=0-4$ were considered. There are many similarities between these two models. In both cases, there are two maxima (one related to g-mode behaviour and the second related to the p-mode behaviour) with a gap in-between them. Both models also cannot explain the highest frequencies in the observed spectrum of HD 163899. The major difference is the higher number of unstable modes in the main-sequence model. It is clear from our calculations, that for a more massive evolved main-sequence models, the behaviour of $\eta$ is similar to the behaviour of $\eta$ for supergiant models. It was not the case for previously studied, less massive models (e.g. \cite[Ostrowski \& Daszy\'nska-Daszkiewicz 2015]{ostrowski2015}). Such similarity makes it impossible to discriminate between evolutionary stages.

For a more reliable comparison, we rotationally split theoretical oscillation spectra and calculated photometric amplitudes in $V$ passband according to the formula of \cite[Daszy\'nska-Daszkiewicz \etal\ (2002)]{dd2002}. The linear theory of stellar pulsations does not provide any information about the intrinsic amplitude, $\epsilon$. We estimated its maximum value, $\epsilon_\mathrm{max} = 0.002$, using the mentioned formula and the highest observed amplitude from the MOST data ($A^{\rm max}_{\rm MOST}\approx 4$ [mmag]). Then, for each mode, we randomly drawn a value for $\epsilon$, from a range $[0,~\epsilon_\mathrm{max}]$. In Fig.\,\ref{fig4} we compare histograms for the observed (the top-left panel) and calculated frequencies for the two main-sequence models differing in mass (the top-right and bottom-left panels) and a supergiant model (the bottom-right panel). We took into account only modes with amplitudes $A_V > 0.3$ mmag, which is the threshold from the MOST data for HD 163899. As expected, the rotational splitting populates the gap between two maxima of $\eta$ in Fig.\,\ref{fig3}. These modes with intermediate frequencies are present in the main-sequence models but their amplitudes are too low in all studied supergiants and that leads to a bimodal distribution of the frequencies. The number of unstable modes is also lower beyond the MS. Generally, the models on the MS much better reproduce the observed frequency distribution of HD 163899 but they are still far from being in a reasonable agreement. For some models we are able to obtain a similar slope of the distribution (the top-right panel in Fig.\,\ref{fig4}) and for the others we got unstable modes with higher frequencies (the bottom-left panel in Fig.\,\ref{fig4}), but we were not able to get both of these features simultaneously. The highest observed frequencies ($\nu > 2.4$) also pose a big challenge.

\section{Conclusions}

For the first time, we obtained basic parameters of HD 163899 using the high-resolution spectra. The new values of $\log T_\mathrm{eff}$ and $\log L/M$ indicate that the star is hotter, more luminous and more massive than than previously thought. On the contrary to the previous results, evolved main-sequence models are located in the error box. According to our initial analyses, they better fit the observed oscillation spectrum than supergiant models and we think HD 163899 is rather in the MS evolutionary stage. The results are preliminary and can be improved as we have problems with explanation of high-frequency peaks of the star and the slope of the observed histogram.

\section*{Acknowledgments}

Based on the archive "HARPS spectra of CoRoT targets", prepared in the framework of the FP7 project n 312844 \textit{SpaceInn - Exploitation of Space Data for Innovative Helio- and Asteroseismology}. This work was financially supported by the Polish NCN grants 2013/09/N/ST9/00611, 2011/01/M/ST9/05914 and 2011/01/B/ST9/05448.

\end{document}